\documentstyle[aps,epsfig,preprint,prl]{revtex}
\tightenlines
\def\grad{\vec\nabla}
\newcommand{\be}{\begin{equation}}
\newcommand{\ee}{\end{equation}}
\newcommand{\ba}{\begin{eqnarray}}
\newcommand{\ea}{\end{eqnarray}}
\begin{document}
\title{Semiclassical description of Stern-Gerlach experiments}
\author{S. Cruz-Barrios$^{1,2}$ and J. G\'omez-Camacho$^1$}
\address{$^1$ Departamento de F\'\i sica At\'omica, Molecular y Nuclear, 
Aptdo. 1065, 41080 Sevilla, Spain}
\address{$^2$ Departamento de F\'\i sica Aplicada 1, E.U.I.T.A. 
Carretera de Utrera, Km.1 Sevilla, Spain } 
\maketitle
\begin{abstract}
The motion of neutral particles with magnetic moments in an inhomogeneous 
magnetic field is described in a semi-classical framework.
The concept of Coherent Internal States is used in the formulation of the
semiclassical approximation from the full quantum mechanical expression.
The classical trajectories are defined only for certain spin states, that 
satisfy the conditions for being Coherent Internal States.
The reliability  of Stern-Gerlach 
experiments to measure spin projections is assessed
in this framework.
\end{abstract}

\newpage
\section{Introduction}

The Stern-Gerlach experiment consists in taking a beam of particles
that have a neutral electric charge, but a finite magnetic moment, and
making them to go through an inhomogeneous magnetic field. The observed
result is that the particles deflect differently depending on the spin
projection along the magnetic field. So, by measuring the deflection, one can
infer the value of the spin projection of the particles along the direction 
of the magnetic field. The Stern-Gerlach experiment 
is the archetype of the measurement of a 
quantum mechanical property. Thus, it is always discussed even in the most 
basic textbooks of quantum mechanics \cite{eis}.
The explanation  that it is usually done for the Stern-Gerlach
experiment is of a semiclassical nature.  The motion of the particles 
is approximated by classical trajectories. In the following, we will present
a brief account of the description of the Stern-Gerlach experiment, as it is
usually done in textbooks.

One starts with a beam of 
particles, in a certain spin state, moving initially along a straight line. 
For the following discussion, 
we will take the y-axis along the direction of the motion of the particles.
The center of the beam corresponds to the coordinates $x=0, z=0$.
The beam enters in a Stern-Gerlach magnet.
Usually, the Stern-Gerlach magnets produce magnetic fields that are independent
of y  and that do not have components in the
y-direction (neglecting border effects). 
The z-axis is chosen along the direction of the magnetic field at the 
center of the beam. So,
\be
\vec B(\vec r) = B_z(x,z) \hat u_z + B_x(x,z) \hat u_x
\ee
where $B_x(0,0)=0$.
The geometry of the Stern-Gerlach magnet is such that $B_z(x,z)$ varies with 
 $z$, but is mostly independent of $x$. Thus, a force appears in 
the z-direction which is proportional
to $dB_z/dz \mu_z$. However, as the divergence of 
the magnetic field has to vanish, then $dB_z/dz=-dB_x/dx$, and so if the
$B_z$ varies with  $z$, then $B_x$ must vary with  $x$. Thus, if we only retain
terms linear in $x$ and $z$, we get
\be
\vec{B}(\vec r) = (B_0 - B_1 z)\hat{u}_z + x B_1 \hat u_x
\ee

The magnetic field interacts with the magnetic moment of the
particle $\mu$, giving rise to an interaction energy given by
\be
E_B(\vec r) = - \vec B(\vec r) \cdot \vec \mu
\ee
This interaction energy depends on the distance. Thus, a  force is generated
which is given by
\be
\vec F = -\grad E_B(\vec r) = \grad B_z(x,z) \mu_z + \grad B_x(x,z) \mu_x
\label{Fzx}
\ee

Thus, a force in the x-direction also appears which would be 
proportional to $dB_x/dx \mu_x$.

>From the expression (\ref{Fzx}), the force acting on a particle with a certain 
spin state can be calculated. If we consider the scattering of particles which
have a given spin projection $m$ along the z-axis, they will suffer a force
that will be calculated as the expectation value
in the z-direction which is proportional to the spin projection. The second
term  in (\ref{Fzx}) will not contribute because the matrix element
$<m|\mu_x|m>$ vanishes. This result is well known, and it is in agreement
with experiment. However, we can consider the case of the scattering of
particles with spin projection $m'$ along the x-axis. If we calculate the
expectation value of $\vec F$ in (\ref{Fzx}), we  obtain a force in the
x-direction, while the force in the z-direction vanishes. This  implies a
deflection of the trajectory in the x-direction, that does not  happen
in the experiment. As any student of quantum mechanics should know, the 
trajectory is split in $2I+1$ trajectories, each one of which is deflected
in the z-direction by a different amount.

In most of the basic textbooks of quantum-mechanics, the term proportional to
$\mu_x$ in (\ref{Fzx}) is simply ignored \cite{eis}.   Other more recent 
books, such as \cite{lev} and \cite{mer}, only consider the spatial variation
of the component $B_z$ of the magnetic field, but they neglect the variation
of $B_x$. The classic book of quantum mechanics
by Messiah \cite{mes} argues that  $\mu_z$ is basically
constant, while $\mu_x$ oscillates around zero. So, in the average of the force
over many oscillations this term would cancel. This argument, although 
plausible, is hardly a firm ground on which to describe the general motion of
particles in inhomogeneous magnetic fields.
The opinion of the authors is that the Stern-Gerlach experiment still requires
a satisfactory description in semiclassical terms. 

In this paper we propose a semiclassical explanation to the Stern-Gerlach 
experiment. It relies on the concept of Coherent Internal States (CIS), that we
introduced in a previous paper \cite{sara}. The conclusion of that paper 
was that when one analyzes the scattering of a particle with
internal degrees of freedom, such as the spin, a single  trajectory  is a
meaningful approximation for the quantum mechanical scattering wave function
only for a certain set of internal states, which we called Coherent Internal
States (CIS). Thus, if the scattered particle has an internal state that 
coincides initially with one of the CIS, then its scattering wavefunction
can be approximated by a single trajectory. If not, the internal state
should be expanded in terms of the CIS, and then the scattering wavefunction
can be approximated by a combination of classical trajectories, one for each 
CIS.

In the case of Stern-Gerlach experiments, we demonstrate in section 2
that the CIS correspond to  states with definite projection along the 
direction of the magnetic field. This direction, called $z'$, may vary
depending on the position of the particle, because the magnetic field is not
homogeneous. So, it does not coincide with the laboratory fixed $z$-axis 
defining the
direction of the magnetic field at the centre of the beam. 
In section 3 we  use the path integral
formalism to describe the trajectories of the CIS, and we find that
they deflect on the $z'$ direction. Considering that the average of $z'$
corresponds to $z$, 
that explains the observed fact that 
the trajectories split in the z-direction, and not in the x-direction,
as (\ref{Fzx}) could suggest.
In section 4 we describe the scattering of a beam of particles with a 
finite size in a Stern-Gerlach magnet. As the axes $z'$ and $z$ do not
coincide,  the deflection of the trajectory is not always consistent with 
the spin projection along the $z$ axis. So, a Stern-Gerlach magnet, understood
as a measurement apparatus to find the spin projection along the $z$ axis,
has a certain probability  of giving a wrong result, which can be evaluated 
in our formalism. 
Section 5 is for the summary and 
conclusions.

\section{Coherent Internal States for  a particle moving in an inhomogeneous
 magnetic field}

It was shown in \cite{sara} that the notion of a classical trajectory is
a useful approach for the quantum mechanical wave function only for
certain selected states, that were called Coherent Internal States (CIS).
These states are the eigenstates of the cross section matrix, 
which are orthogonal and form a basis of the space of the initial internal 
states.
To find the CIS, as it is was shown in \cite{sara}, the 
following iterative procedure should be followed:
\begin{itemize}
\item Solve the classical scattering problem for the uncoupled 
hamiltonian $H_0$ and obtain the 
evolution operator along the classical trajectory. 
\item Consider small desviations from the classical 
trajectory. Evalute the operator $\Delta$, defined in \cite{2}, which
describes the dependence of the cross section matrix on the initial state.
\item Obtain the CIS $|n>$ diagonalizing the cross section matrix, which is
equivalent to diagonalize the operator $\Delta$. 
\item Evaluate the  classical trajectories ${\bf r}_n(t)$ for each CIS, 
the evolution operators $U_n(t,t_0)$, and the final states $|\tilde{n}>$.
If the final states are orthogonal, then the calculated cross section matrix 
will be diagonal, and the self-consistency would have been achieved. 
If not, the CIS should be  recalculated as the eigenstates of the cross section
matrix and the procedure should be followed until self-consistency is achieved.
\end{itemize}

Let us consider a neutral particle of mass $M$ that moves in the y-direction 
with velocity $v_y$ and which has an 
initial position characterized by the coordinates $(x_0, z_0)$. Note that
in strict quantum mechanical terms, we can take wave 
packets sufficiently localized around $(x_0, z_0)$, which would have momentum 
dispersions much smaller than $M v_y$.
This particle, that has a spin $I$ and a magnetic moment $\vec \mu = \mu_0
\vec I$, enters in a magnetic field.
The time evolution of this particle moving in a magnetic field is given by the 
evolution operator ${U}(t_f,t_0) = \exp \{-i/\hbar \int_{t_0}^{t_f} {H}
dt \} $, where the Hamiltonian  $ H $ of the system is written as
\be
 H = \frac{p^2}{2M}  - \vec{\mu} \cdot \vec{B}
\label{ham}
\ee 

The first term is the kinetic energy of the particle and the second term
is the potential interaction of the particle with the magnetic field.  
The  magnetic field has  $x, z$ components so that, 
\be
\vec{\mu}\cdot \vec{B} = \mu_0[(B_0 - B_1 z){I}_z + x B_1 {I}_x]
\ee 

We will 
define new axes $(x', z')$ so that the magnetic field
$\vec B(x_0,z_0)$ is directed along $z'$. The y axis is unaffected.
The angle  $\beta(x_0,z_0) $ that  
generates the rotation  is given by
\be
\tan(\beta(x_0,z_0)) = x_0 B_1 / (B_0 - z_0 B_1)
\ee

The angular momentum operators in the new coordinate system are
\ba
{I'}_z = {I}_z \cos \beta + {I}_x \sin \beta \nonumber \\
{I'}_x = -{I}_z \sin \beta + {I}_x \cos \beta 
\ea \label{irot}

The  interaction term with the magnetic field is given by, 
\be
\vec{\mu}\cdot \vec{B} = \mu_0 B'_0 {I'}_z + \mu_0 B_1 [-(z-z_0){I}_z+
(x-x_0)I_x]
\ee 
with 
\be B'_0 = \sqrt{ (B_0 - B_1 z_0)^2 + (B_1 x_0)^2} \ee

The Hamiltonian (\ref{ham}) can be written  as:
\ba 
H &=& H_0 + V \\ 
H_0 &=& \frac{p^2}{2M} - \mu_0 B'_0 {I'}_z \\ 
V &=&  \mu_0 B_1 [-(z-z_0)I_z+
(x-x_0)I_x]
\ea
For the case that we are considering, the classical trayectories for $ H_0$ 
are just straight lines,  given by the expressions:
\be
\vec{r}(t) =  \vec{r}_0 + \frac{ \vec{p}_0}{M}t  = 
\vec{r}_f - (t_f - t) \frac{\vec{p}_0}{M} \label{ecr}
\ee
We consider the trajectory of a particle that is initially in the position 
$x=x_0$, $z=z_0$, and
moves in the $y$ direction. The evolution operator for $H_0$   is: 
\be 
{U}_0(t,t_0) = \exp \{ \frac{i}{\hbar}[ (t-t_0) \mu_0 B'_0 \hat{I'}_z ]\}.
\ee
In a basis $\{|m>\}$ of eigenstates of $I'_z$, 
the matrix elements of the evolution operator
 associated with the classical trajectory are diagonal:
\be
<m_f|T^0|m_i> = < m_f |U_0(t_f,t_0) |m_i> = \delta_{m_f,m_i} \exp 
\{- i \omega_0(t_f - t_0)m_i \}
\ee 
with $\omega_0 = \mu_0 B'_0 /\hbar $. 

When the effect of small deviations from the classical trajectory in the path
integral formalism are considered, the expression for the scattering amplitude
becomes
\ba
<m_f|T|m_i> & = & <m_f|T^0|m_i> + <m_f|\delta T|m_i> =  \nonumber \\ 
            &  & <m_f|T^0|m_i> + <m_f|T^1|m_i> + <m_f|T'|m_i>
\ea
The  correction terms are given  by \cite{2}:
\ba
<m_f|T^1|m_i> & = &  -\frac{1}{2} \int^{t_f}_{t_0} dt 
<m_f|U_0(t_f,t) \left( \frac{\partial}{\partial \vec{r}_0} 
\right)_{\vec{p}_0} 
\left( \frac{\partial V(t)}{\partial\vec{p}_0} \right)_{\vec{r}_f} 
{U}_0(t,t_0)|m_i>  \label{t1} \\
<m_f|T'|m_i>& = & \frac{i}{2 \hbar} \int^{t_f}_{t_0} dt \int_{t_0}^t dt'
<m_f|{U}_0(t_f,t) \left[ \left (\frac{ \partial V(t)}{ \partial \vec{p}_0} 
\right)_{\vec{r}_f} U_0(t,t') \left( \frac{ \partial V(t')}
{ \partial \vec{r}_0} \right)_{\vec{p}_0} \right. \nonumber \\
 &  &  -   \left. \left( \frac{\partial V(t)}{\partial \vec{r}_0} 
\right)_{\vec{p}_0} U_0(t,t') 
\left( \frac{\partial V(t')}{\partial \vec{p}_0} 
\right)_{\vec{r}_0}\right] U_0(t',t_0) |m_i>  \label{t'}
\ea
The derivatives w.r.t. $\vec{p}_0$ can be expressed in terms of derivatives
w.r.t. $\vec r$ using eq. (\ref{ecr}).
The term $<m_f|T^1|m_i>$ vanishes because $ \nabla^2 \vec{B} = 0$ and so
$\nabla^2 V(t) =0 $. The  term $T'_{m'm}$ can be calculated in a 
straightforward way resulting:
\ba
<m_f|T'|m_i> = \left( \mu_0 B_1 \right)^2 \frac{i}{2 \hbar M} 
\int^{t_f}_{t_0} dt
\int^{t}_{t_0}dt'(t - t') & 
<m_f |\left[ U_0(t_f,t) I_x  U_0(t,t') 
I_x U_0(t', t_0) \right. \nonumber \\ 
& \left.  +  U_0(t_f,t)I_z U_0(t,t') 
I_z U_0(t',t_0) \right] |m_i> 
\ea
This expression can be written in terms of the rotated angular momentum 
operators $I'_x, I'_z$:
\ba
<m_f|T'|m_i> =  \left( \mu_0 B_1 \right)^2 \frac{i}{2 \hbar M} 
\int^{t_f}_{t_0} dt
\int^{t}_{t_0}dt'(t - t') & 
 <m_f |\left[ U_0(t_f,t) I'_x U_0(t,t') 
I'_x U_0(t',t_0) \right. \nonumber \\
 &   \left. + U_0(t_f,t) I'_z U_0(t,t') 
I'_z U_0(t',t_0) \right] |m_i> 
\ea

Then, the first-order correction $\Delta_{m'm}$ defined as;
\be
<m'_i|\Delta|m_i> = \sum_{m_f}[ <m_f|T^0|m'_i>^* <m_f|\delta T|m_i> + 
 <m_f|T^0|m_i> <m_f|\delta T|m'_i>^* ]  
\ee
is given by:
\ba
<m'_i|\Delta|m_i>  = 
\left( \mu_0 B_1 \right)^2 \frac{i}{2 \hbar M} \int^{t_f}_{t_0} dt
 \int^{t}_{t_0}dt' (t-t')    
&  <m'_i | \left[U_0(t_0,t){I'}_x U_0(t,t')
{I'}_x \hat{U}_0(t',t_0) +   \right. \nonumber \\  
 &   \left. U_0(t_0,t) {I'}_z U_0(t,t')
 {I'}_z U_0(t',t_0) \right] |m_i> + h.c.
\label{del}
\ea
The non-diagonal matrix elements of (23) vanish. 
The diagonal matrix elements are given by: 
\be
<m'_i|\Delta|m_i> =  \delta_{m'_i, m_i}
 \frac{2 m_0}{M \hbar} \frac{B_1 \mu_0}{\omega_0^3}
\left[ 2 - 2 \cos(\omega_0 (t_f-t_0)) -
(\omega_0 (t_f - t_0) \sin  \omega_0 (t_f -t_0)) \right] 
\ee
 
This result shows that the states $|m_i>$, with a definite spin projection 
along the $z'$ axis, are the eigenstates of $\Delta$, and thus they are  
our initial choice for  CIS.  As we will see in the next section, these
states are not modified as the particle moves along the classical trajectory,
and so they are the Coherent Internal States of our problem.

The original states $|m;L>$, which have a definite spin 
projection $m$ along the laboratory fixed $z$-axis, are not 
CIS. To describe the evolution of these states, they should be expanded in 
terms of the CIS, by means of the expression: 
\be
|m;L> = \sum_{m'}d^I_{m',m}( \beta(x_0,z_0))|m'>
\ee

\section{Semiclassical description of the scattering for Coherent Internal
States}

We will now evaluate the classical trajectory 
for each CIS, by making the stationary phase approximation on the 
matrix elements of the exact propagator on the calculated CIS.

The exact propagator of the system between the CIS is  given as a path
integral extended to all possible trajectories by:
\be
<m(f)|K|m> = \int D[{\vec r(t)}] 
\exp \{ \frac{i}{\hbar} {S}_{eff}(\vec r(t)) \} 
\label{propa}
\ee
where 
\be 
{S}_{eff} = {\vec r}_0 \cdot{\vec p}_0 + {S}_0(\vec r(t)) 
- i\hbar\ln <m(f)|U_B (\vec r(t); t_f, t_0) |m> 
\ee
is the effective action, $S_0(\vec r(T))$ is the action corresponding to the 
kinetic energy, and
\be
 U_B (\vec r(t); t_f, t_0) = 
\exp \{ \frac{i}{\hbar} \int_{t_0}^{t_f} dt  
\vec{\mu} \cdot \vec{B}(\vec r(t)) \}
\label{ope}
\ee
is the evolution operator of a particle in magnetic field along the trajectory
$\vec r(t)$.
In this expression, $|m>$ is one of the CIS, and $|m(f)>$ is the final
state, defined by
\be
|m(f)> = U_B(\vec r_m(t);t_f,t_0) |m> 
\ee
The classical trajectory $\vec r_m(t)$ is obtained
 imposing the stationary phase condition in (\ref{propa}),  
\be 
\frac{ \delta S_{eff}}{\delta \vec r(t)} \arrowvert_{\vec r(t) 
= \vec r_m(t)} = 0 
\ee
This  leads in a strightforwad way to:
\ba
{d \vec r_m(t) \over dt}\arrowvert_{t=t_0} &=& { \vec{p}_0 \over M} 
\nonumber \\
{d^2  \vec r_m(t) \over dt^2} & = & - \frac{1}{M} 
< m(t)| \frac{ \partial}{ \partial \vec{r}}
(\vec{ \mu} \cdot \vec{B}(\vec r)) |_{\vec r = \vec r_m(t)} | m(t)>
 \label{cla}
\ea
where the state at the instant $t$ is given by 
$|m(t)> = U_B(\vec r_m(t);t,t_0) |m>$. These  equations describe 
the motion of a classical particle with a magnetic moment $\vec \mu(t)
= <m(t)|\vec \mu|m(t)>$ moving in the inhomogeneous magnetic field 
$ \vec{B}(\vec r)$. However, it should be stressed that this interpretation is
only meaningful for internal states $|m(t)>$ that evolve from a CIS state
$|m>$ at $t=t_0$.

We observe that in the $y$ direction we have a constant  motion given by
$ y_f = y_0 +(t_f - t_0) \frac{p_{0}}{M} $ and in the $x, z$ directions the 
motion is accelerated and the force is  proportional to $B_1$ because:
\be 
<m(t)|\frac{ \partial}{ \partial \vec{r}} (\vec{ \mu} \cdot \vec{B}(\vec r))
|m(t)> 
=  - \mu_0 B_1(<m(t)|I_z|m(t)> \hat u_z - <m(t)|I_x|m(t)>  \hat u_x)
\ee

Known the force, the specific nature of the classical solution in the $x z $ 
directions depend  enterily of the evolution operator  $ U_B$ and the 
$|m>$ state. Some considerations must be done to solve the 
classical equations in the $x, z $ directions.
It should be noticed that the trajectory of the particle, that initially is
on $(x_0, z_0)$, is always along the line that joins this point with the
point with coordinates $(x=0, z=B_0/B_1)$. Along this line, the direction of
the magnetic field is fixed, although its magnitude will vary. The state
$|m(t)>$ is given by the initial CIS $|m(t)>$ times a phase factor.
The angle
$\beta$ between the axes $z$ and $z'$ is constant. Thus, we have
\be
<m(t)|I_z|m(t)> = \cos(\beta(x_0,z_0)) \quad ; \quad <m(t)|I_x|m(t)>
 =  \sin(\beta(x_0,z_0)) 
\ee

So, the trajectory is uniformly accelerated along the $x$ and $y$ directions,
so that, when they leave the magnetic field at a time $t_f$, the coordinates
have changed to
\ba
z_f &=& z_0 - {\mu_0 B_1 m \over 2 M} (t_f-t_0)^2 \cos(\beta(x_0,z_0)) \\
x_f &=& x_0 + {\mu_0 B_1 m \over 2 M} (t_f-t_0)^2 \sin(\beta(x_0,z_0)) 
\ea 
When the particle leaves the magnetic field, the force vanishes.
However, as it has acquired a certain velocity, 
the values of these coordinates at a later time $t_d$, in which they are
detected, is given by 
\ba
z_d &=& z_0 - b m \cos(\beta(x_0,z_0)) \\
x_d &=& x_0 + b m \sin(\beta(x_0,z_0)) 
\ea 
where $b = \mu_0 B_1  (t_f-t_0)(2t_d - t_f-t_0)/(2M)$, measures the spacial
separation 
between the different magnetic substates. Note that, by making $t_d$ 
sufficiently large, we can obtain reasonable values of $b$, even if the
gradient of the magnetic field $B_1$ is small, and the length of the magnet,
which is $L=(t_f-t_0) v_y$, is small.

\section{Reliability of Stern-Gerlach experiments to measure spin 
projections}

We will consider a beam of particles with a finite extension. Thus, the center
of the beam will be placed at $x=0, z=0$, but it will have a probability
density $P(x,z)$ of being in the position $(x,z)$. We will assume, to
make the calculations simpler, that this probability density is gaussian:
\be
P(x,z) = {1 \over \pi a^2} \exp( - {x^2+z^2\over a^2}) \label{dist}
\ee
so, $a$ is a measurement of the size of the beam. Besides, we will
assume that all the velocity of all the particles goes essentially along the 
$y$ direction, with a velocity $v_y$. That means that the $x$ and $z$ 
components of the velocity must be very small so that $v_z (t_d-t_0) \ll b$.
The uncertainty principle implies that $M v_z > \hbar/a$, and so one gets,
for $(t_d-t_0)\gg(t_f-t_0)$,
the condition
\be
\hbar \ll B_1 a (t_f-t_0) = {B_1 a L \over v_y}.      \label{cond1}
\ee
This condition implies that the gradient of the magnetic field  cannot 
be arbitrarily small. However, we will see that it cannot
be too large either.

Let us consider the particle which is initially  in the position $(x,z)$. 
The magnetic field that it sees is given by
\be
\vec B(x,z) = (B_0 - z B_1) \hat u_z + x B_1 \hat u_x
\ee
Thus, the direction of the magnetic field does not go along the z-direction.
The angle $\beta(x,z)$ between $\vec B(x,z)$ and $\hat u_z$ is given by
\be 
\tan \beta (x,z) =  x B_1 /(B_0 - z B_1)
\ee
Note that the states  $|m;L>$ with spin projection $m$ along the 
laboratory fixed 
$z$-axis are no longer
the CIS. For a given value of $(x,z)$, the CIS are states $|m'>$ which have
a given spin projection $m'$ along the axis $z'$ which is paralell to  
$\vec B(x,z)$. Then, the state $|m;L>$ will not give rise to a unique 
trajectory,
but to $2I+1$ trajectories which are characterized by the states
$|m'>$. The probability that the state $|m;L>$ follows the trajectory of 
$|m'>$ is simply  
\be
p(m,m') = |<m;L|m'>|^2 = |d^I_{m,m'}(\beta(x,z))|^2.
\ee

We have considered the case of a spin-1 particle. It has three possible spin
projections along the laboratory fixed $z$-axis, corresponding to the states
$|m;L>$, for $m=-1,0,+1$. When a beam of particles with these spin states
go through a Stern-Gerlach magnet, they will be deflected according to the
value of the spin projection $m'$ along the $z'$-axis. The probability that
the deflection of the particles is determined by the spin projection along
the laboratory fixed $z$-axis, is given by 
the expectation value of $p(m,m)$ averaged over the beam distribution,
is shown in figure 1. The values $<p(1,1)>$ and $<p(0,0)>$ are represented
as a function of the dimensionless parameter $aB_1/B_0$. We see that only 
for values $aB_1/B_0 \ll 1$, the values of $<p(1,1)>$ and $<p(0,0)>$ 
tend to one. This indicates that the deflection of the particles will be 
determined by the spin projection in the ``laboratory'' fixed axis $z$ only
if  $aB_1/B_0 \ll 1$. Thus, this is the condition for a Stern-Gerlach 
experiment to be a measurement of the spin projection.

\begin{center}
\begin{figure}[hbt]
\mbox{\epsfig{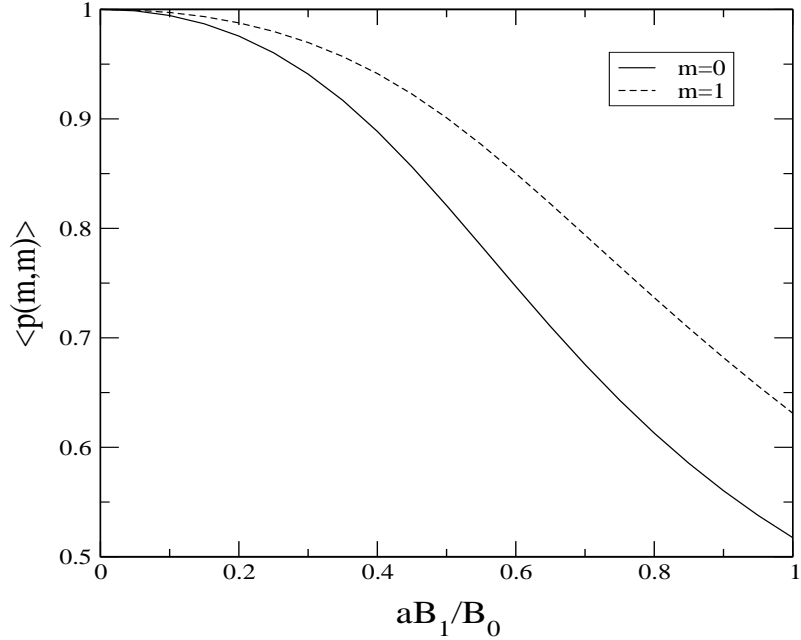}}
\vspace{1cm}
\caption{Probability that a particle with spin 1 and projection $m$ along
the laboratory fixed $z$-axis is deflected according its spin projection, 
as a function of the 
parameter $aB_1/B_0$, for $m=0$ and $m=1$. The case $m=-1$ coincides with 
$m=1$}  
\end{figure}
\end{center}

If we will consider the case in which $a B_1 \ll B_0$, then, the angle 
$\beta(x,z)$ is small, and the expression above can be expressed as
\ba
p(m,m') &=& \delta_{m,m'}\left(1-{I(I+1)-{m}^2 \over 2}
                   \beta(x,z)^2\right) \\
        &+&\delta_{{m}+1,m'}\left({I(I+1)-{m}({m}+1) \over 4}
                   \beta(x,z)^2\right) \\
        &+&\delta_{{m}-1,m'}\left({I(I+1)-{m}({m}-1) \over 4}
                  \beta(x,z)^2\right) 
\ea
The Stern-Gerlach experiment is used to measure spin projections along the
z-axis. Thus, if we observe a deflection corresponding to the state $m'$,  
we would conclude that the spin projection along the laboratory fixed
$z$-axis was $m'$, instead of $m$.
Thus, the probability that the measurement gives the incorrect result is
\be
p_e({m}) = \left({I(I+1)-{m}^2 \over 2}\beta(x,z)^2\right) 
\ee
If we average this probability over the beam probability density, we obtain
\be
<p_e({m})> =\left({I(I+1)-{m}^2 \over 2}\right) {a^2 B_1^2 
                       \over 2 B_0^2}
\ee

Thus, we see that the crucial condition for the Stern-Gerlach experiment to
be useful in order to measure spin projections is that the magnetic field
$B_0$ should be much larger than the product of the gradient of this field
multiplied by the size of the beam. In other words, the relative change of 
the magnetic field within the finite extension of the beam sould be very small.
Thus, we can write
\be
B_0 \gg B_1 a.      \label{cond2}
\ee
Putting together the conditions in eqs.(\ref{cond1},\ref{cond2}), we have
\be
{B_0 (t_f-t_0) \over \hbar} \gg {B_0 \over B_1 a} \gg 1. \label{cond12}
\ee
However, the first term is just the precession angle of the magnetic moment
operator about the z-axis. This angle has to be very large, compared to 1,
as a neccesary condition for eq. (\ref{cond12}) to be valid. In this sense,
our results are in agreement with the argument of Messiah \cite{mes}, 
which points to the fact that the x-component of the magnetic moment 
oscillates around zero, and then it can be ignored. However, we find that this
argument is not sufficient. The gradient of the magnetic field has to be such 
that $B_0 \over a B_1$ is much smaller than the precession angle, and much 
larger than one.  

We have performed a simulation of a beam of 1000
particles, with spin 1 and projections along the laboratory fixed $z$-axis 
$m=-1,0,+1$,
and  a  probability density of having initial $(x,z)$ values 
given by equation (\ref{dist}).  The particles go through  an inhomogeneous 
magnetic field, and are deflected. The values of the magnetic field, 
its gradient, the length of the magnet, and the distance of the detectors are
taken so that the parameter $b$, wich determines the amount of the deflection, 
is given as $b=4a$, in terms of the initial size of the beam. 
The relation of the
magnetic field and its gradient is given by $B_1a/B_0=0.25$.
In figure 2 we present the results of the simulation, presenting the final 
values of the $x$ and $z$ coordinates, in units of $a$, for different values 
of $m$. We find that, in general, most of the particles suffer deflections
along the z axis consistent with their spin projection, but there are a few
cases (corresponding to about 3\% for $m=0$, and 1.5\% for $m=\pm1$), in which
this is not the case. This is what we expect from figure 1. The other effect
that one sees in figure 2 is a focusing effect for the particles with spin
projection $m=-1$. That is related to the fact that the deflection occurs
along lines that cross in the point with $(x=0, z=B_0/B_1)$.  The states with
$m'=-1$ tend to get close to this point, and so they focus, while the states
with $m'=1$ tend to separate from it, and so they de-focus.
    
\begin{center}
\begin{figure}
\mbox{\epsfig{file=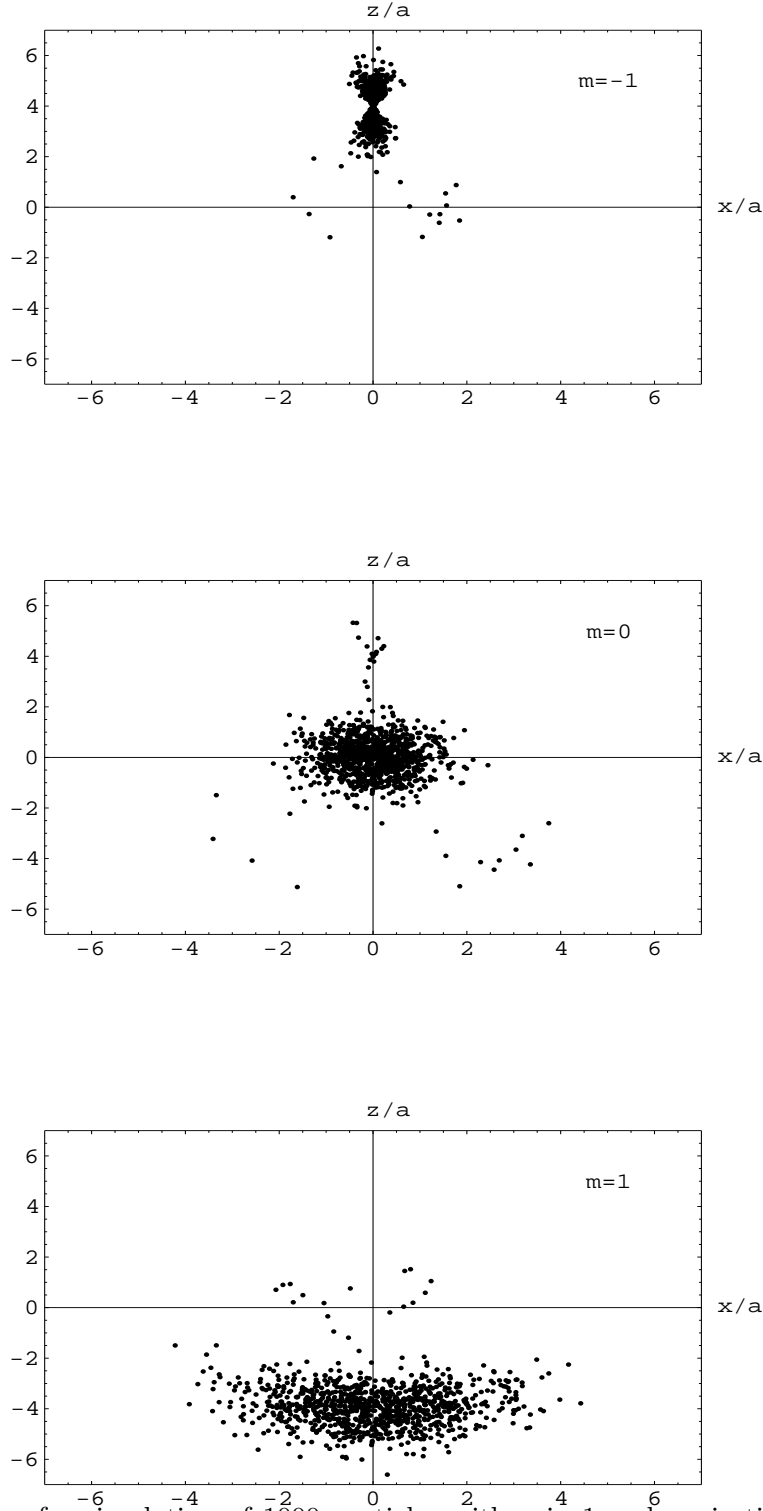, height=20cm, width=10cm}}
\caption{Results of a simulation of 1000 particles with spin 1 and projections
$m=-1,0,+1$ which go through an inhomogeneous magnetic field described by
$b=4a$ and $B_1a/B_0=0.25$. The final position of the particles in the $(x,z)$
plane is represented in units of the initial beam size $a$.}
\end{figure}
\end{center}

In this work we have focussed in the application of Stern-Gerlach magnets as
a measurement apparatus to determine the spin projection of individual atoms,
assuming that the spin and the magnetic moment is previously known. 
We find that these experiments are
not completely reliable, because the deflection is not uniquely determined by 
the spin projection. However, Stern-Gerlach experiments may also be used
to measure the magnetic moment. In this case, one should measure the separation
of the piles of particles coming from an initially unpolarized beam. 
Our calculations could be useful for this purpose, because they not only give
the separation, but also give the shape of the different piles. 

\section{Summary and conclusions}

We describe the motion of a particle with spin in an inhomogeneous magnetic 
field, in a semiclassical approach. We make use of the fact that
the  classical trajectories are only a meaningful approach to the quantum
mechanical scattering wavefunction for certain states of the internal
variables, that are called Coherent Internal States. The Coherent Internal
States are obtained initially as the eigenstates of the cross section matrix. 
Each one of these states has a trajectory that describes the time dependence 
of the
coordinate, and an evolution operator that describes the time dependence 
 of the internal state. The trajectory and the evolution operator are 
related self-consistently, because the classical force that
defines the trajectory is related to expectation value of the coupling
potential on the internal state, while the evolution operator is related 
to the coupling potential evaluated along the trajectory.

We have considered the case of a particle with a given magnetic moment,
moving initially in the y-direction within a magnetic field that depends 
linearly on 
the coordinates x and z. The Coherent Internal States correspond 
initially to definite projections of the spin along the direction of the 
magnetic field 
evaluated at the initial point of the trajectory. The trajectory corresponding
to a given spin proyection will be deflected due to the gradient in the 
magnetic field. The magnitude of the magnetic field observed by each particle 
may change as a result of the deflection of the trajectory.
However, the direction of
the magnetic field remains constant. Thus, the Coherent Internal State does not
change as the particle moves along its trajectory.

We have considered an ensambe of particles, all moving in the y-direction,
with initial x and z coordinates following a gaussian probability distribution.
This illustrates a realistic situation for a beam of particles of finite size
entering a Stern-Gerlach magnet. We have evaluated the trajectories followed by
these particles, considering explicitly that the Coherent Internal States are
different for different particles of the ensamble, because they depend on
the initial values of x and z. We find that not all the particles having a 
given spin projection along the z-axis suffer the same deflection. 
This indicates that even
an idealized Stern-Gerlach experiment has a finite probability of giving
the wrong result as a measurement apparatus of the spin projection. The
probability of error depends on the relative dispersion of the values of 
magnetic field within the beam size.

\bigskip

{\bf Acknowledgements:} This work has been partially supported by the
spanish CICyT, project PB98-1111

\end{document}